\documentclass[letterpaper, 10 pt, conference]{ieeeconf}  

\IEEEoverridecommandlockouts                              

\usepackage{graphicx} 
\usepackage{amsmath} 
\usepackage{amssymb}  
\usepackage{footnote}
\usepackage{lipsum} 
\usepackage{multirow}
\usepackage{multicol}
\usepackage{caption}
\usepackage{subcaption}
\usepackage{xcolor}

\title{\LARGE \bf
Real-time Audio Video Enhancement \\with a Microphone Array and Headphones
}

\author{Jacob Kealey, Anthony Gosselin, \'Etienne Deshaies-Samson, Francis Cardinal, \\F\'elix Ducharme-Turcotte, Olivier Bergeron, Am\'elie Rioux-Joyal, J\'er\'emy B\'elec, Fran\c{c}ois Grondin 
\thanks{This work was supported by NSERC-CREATE/CoRoM and a NSERC Discovery Grant. Authors are with the Department of Electrical Engineering and Computer Engineering, Universit\'e de Sherbrooke.}%
}

\begin{document}

\maketitle
\thispagestyle{empty}
\pagestyle{empty}

\begin{abstract}
This paper presents a complete hardware and software pipeline for real-time speech enhancement in noisy and reverberant conditions.
The device consists of a microphone array and a camera mounted on eyeglasses, connected to an embedded system that enhances speech and plays back the audio in headphones, with a latency of maximum 120 msec.
The proposed approach relies on face detection, tracking and verification to enhance the speech of a target speaker using a beamformer and a postfiltering neural network.
Results demonstrate the feasibility of the approach, and opens the door to the exploration and validation of a wide range of beamformer and speech enhancement methods for real-time speech enhancement.
\end{abstract}

\section{INTRODUCTION}

Robot audition aims to provide robots with hearing capabilities similar or superior to humans.
This often involves speech enhancement in noisy and reverberant conditions, with online processing on low-cost hardware and little latency.
Frameworks such as ManyEars \cite{grondin2013manyears}, ODAS \cite{grondin2019odas} and HARK \cite{nakadai2010design} have been proposed in the past to provide robots with the ability to localize, track and separate sound sources using microphone arrays.
On the other hand, listeners with hearing loss experience difficulty listening to a specific conversation in a noisy environment (with babble noise for instance) \cite{brons2014effects}.
This paper proposes to adapt some of the existing methods used in robot audition to provide a better hearing experience for impaired listeners.

Beamforming combines multiple audio channels together to enhance a specific sound of interest, based on its spatial properties.
Delay-and-sum beamforming can be easily performed in the time or frequency domains if the target sound source direction of arrival (DoA) and the microphone array geometry are known \cite{perrot2021so}.
It provides a low-complexity and low-latency solution, but is subject to leakage from competing sources and spectral coloration due to early reflections.
To deal with competing sound sources, geometric sound separation (GSS) minimizes online a cost function to enhance the sound source of interest and null the interfering sources \cite{parra2002geometric, valin2004enhanced}.
The method however requires the DoA of all sound sources (both target and interfering), and usually converges slowly to a solution, especially if sound sources are moving in time.
With GSS, dealing with high reverberant environment also becomes challenging as the cost function tries to simultaneously satisfy the freefield propagation assumption and ensure statistical independence between the separated sources. 
On the other hand, minimum variance distortionless response (MVDR) beamformer aims to provide a constant gain in the direction of the target source, while reducing any competing sources coming from other directions, such that the DoA of interfering sources can be unknown \cite{habets2009new}.
This method is efficient but requires an estimation of the spatial covariance matrices (SCMs) for both the target and noise.
The SCMs are usually obtained using time-frequency masks estimated using a deep neural network \cite{erdogan2016improved}, which statistics usually need to be computed over a few seconds of audio.
This constraint makes MVDR less suitable for real-time scenarios.
Generalized Eigenvalue Decomposition (GEV) beamforming can be more robust to inaccuracies while estimating SCMs \cite{heymann2015blstm, grondin2020gev}, but it also relies on statistics over a few seconds of audio, and involves online eigenvector decomposition, which remains challenging on low-cost hardware.

End-to-end multichannel neural network based approaches have also been proposed to perform sound source separation.
Deep clustering can exploit the spatial and spectral features of speech to separate individual speakers using clustering \cite{wang2018multi}.
This method however works for speech signals only, and requires some buffering to perform statistically meaningful clustering.
Another approach consists in adapting the Conv-TasNet model \cite{luo2019conv} to deal with multiple channels \cite{lee2021inter}.
While it provides high quality speech separation, this strategy involves more than one million parameters, and ignores the target DoA to listen only to a specific speaker.

Recently, work has been done with an augmented reality headset to perform robust online transcription of reverberant and noisy speech of a speaker \cite{sekiguchi2022direction}.
In this case, automatic speech recognition (ASR) is the end goal, which allows for a latency of a few seconds.
However, for real-time speech enhancement, latency should be in the order of tens of milliseconds, to avoid unpleasant delay between the lips motion and the fedback enhanced speech signal.
In fact, studies show that in general speech and lips asynchronicity should not exceed a latency of 140 msec \cite{summerfield1992lipreading}.

In this paper, we demonstrate the feasibility of designing a hardware and software pipeline to perform real-time speech enhancement with a set of headphones, and a microphone array with a camera installed on the frame of eyeglasses. 
The hardware and software are specifically designed to minimize the latency, and the algorithms are chosen to allow online processing on a NVIDIA Jetson Xavier NX embedded computer.
Section \ref{sec:proposed_system} describes the algorithms and the custom-made hardware of the proposed system.
Section \ref{sec:experimental_setup} introduces the experimental setup used to validate the performances of the proposed approach, and section \ref{sec:results} discusses the results.
In section \ref{sec:conclusions}, we conclude this paper and discuss possible future work.

\section{PROPOSED SYSTEM}
\label{sec:proposed_system}

\begin{figure*}
    \centering
    \vspace{10pt}
    \includegraphics[width=\linewidth]{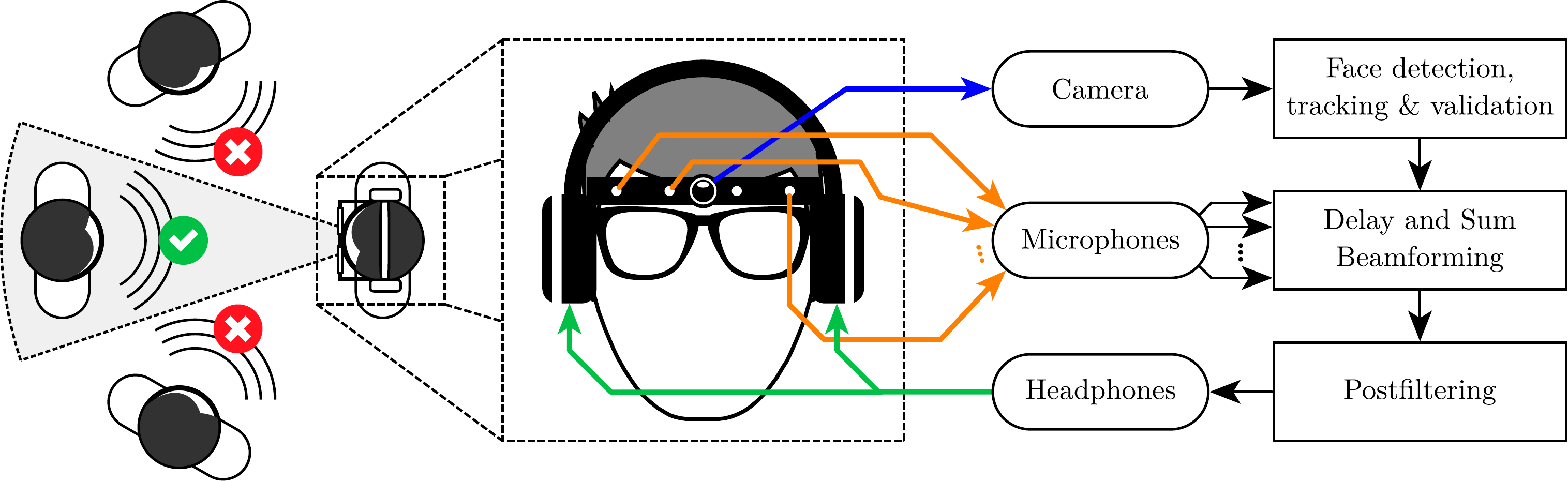}
    \caption{Overview of the proposed system. The camera on the eyeglasses capture the visual scene, and localizes the target face. The position is converted in a direction of arrival that is used by the delay and sum beamformer to enhance the target speech. A postfiltering step is then applied to improve the speech quality and this is returned to the headphones as an audio feedback.}
    \label{fig:overview}
\end{figure*}

Figure \ref{fig:overview} shows an overview of the system.
The user wears headphones and eyeglasses, which are equipped with a 8-microphone array (2 on each side and 4 in the front) and a small camera facing the scene.
The camera provides a video stream to perform face detection and track the target, which can be selected using a web application running on a smartphone.

\subsection{Face detection, tracking and verification}

The camera mounted on the eyeglasses look at the speakers from the listener point of view and feeds an image used to estimate the position of the target with respect to the listener.
This is achieved in three steps inspired from the DVT method \cite{kalal2010face, zhang2018long}: 1) face detection, 2) tracking and 3) verification.

Face detection uses the YOLOv5 model to scan the optical image and return one or multiple bounding boxes around faces \cite{xu2021effective}.
Detection can be performed at a low refresh rate (one frame per second) to detect new speakers who join the visual scene.
Tracking is performed using Kernelized Correlation Filters (KCF) \cite{henriques2014high}, as this method provides robust result and allows high-speed tracking (with a rate of 20 frames per second for this application), which is needed here for an embedded system with limited computing resources.
The listener chooses the target face to listen to using a web application that runs on a smartphone.
Face verification can compare the target speaker face to the tracked faces using a ResNet18 trained for the ReSORT algorithm \cite{tran2021resort}, each 1.5 sec, to ensure object permanence (in case of temporary face occlusion for instance).
The tracked position of the face of the target speech on the image is forwarded to the beamformer module at rate of $4$ frame per second to enhance the target speech.

\subsection{Delay and Sum Beamforming}

The audio signals are sampled at a rate of $16000$ samples/sec.
Each signal is represented by $x_m[n]$ in the time domain, where $m \in \{1, 2, \dots, M\}$ stands for the microphone index (for a total of $M \in \mathbb{N}$ microphones) and $n \in \mathbb{N}$ for the time index.
These input signals are transformed with Short-Time Fourier Transform (STFT) with a frame size of $N \in \mathbb{N}$, a hop size of $\Delta N \in \mathbb{N}$ samples and a sine window, and denoted as $X_m[l,k] \in \mathbb{C}$, where $l \in \mathbb{N}$ stands for the frame index and $k \in \{0, 1, \dots, N/2\}$ for the frequency bin index.
In the proposed system, the frame size is $N=512$ and the hop size is $\Delta N = 256$.
The sine window (or root-Hann) is power complementary, which allows analysis and synthesis with an overlap of $50\%$.

The time difference of arrival (TDoA) between each microphone $m$ and the origin (positioned at the center of mass of the microphone array) is denoted by $\tau_m[l] \in \mathcal{T}$, where $\mathcal{T} = [-\tau_{max}, +\tau_{max}]$, and $\tau_{max}$ represents the maximum TDoA value, which depends on the distance between the microphones, the sample rate and the speed of sound.
The delay-and-sum beamforming result is obtained in the frequency domain as follows:
\begin{equation}
    Y[l,k] = \sum_{m=1}^{M}{X_m[l,k] \exp{\left(j\frac{2\pi k \tau_m[l]}{N}\right)}},
    \label{eq:Y}
\end{equation}
where $j=\sqrt{-1}$.
While TDoAs can be estimated based on the audio signals only \cite{anguera2007acoustic,grondin2019multiple}, it is desirable to use the visual feedback to identify the target of interest.
This can be achieved by using the target speaker's lips position obtained from the optical image, and map the pixel $xy$-coordinate to TDoAs.
A function maps each pixel denoted by the coordinate $(u[l],v[l])$, where $u[l] \in \mathcal{U}$, with $\mathcal{U} = \{1, 2, \dots, U\}$ and $v[l] \in \mathcal{V}$, with $\mathcal{V} = \{1, 2, \dots, V\}$, for an image with a width of $U \in \mathbb{N}$ pixels and a height of $V \in \mathbb{N}$ pixels.
This is achieved using polynomial regression from a simple calibration step, as in \cite{grondin2020audio}:
\begin{equation}
    f: \mathcal{U} \times \mathcal{V} \mapsto \mathcal{T}^M.
\end{equation}
With this mapping, the TDoAs are obtained as follows:
\begin{equation}
    \{\tau_1[l], \tau_2[l], \dots, \tau_M[l]\} = f((u[l],v[l])),
\end{equation}
and then substituted in (\ref{eq:Y}).
It is possible for the target speaker to move, or the listener to turn its head: this is handled as the pixel coordinates change over time for each frame $l$, which updates the TDoAs accordingly. 

\subsection{Postfiltering}

While beamforming increases the gain of the target source, it is also subject to interference leaking from other directions.
We propose a postfiltering final step to perform time-frequency filtering.
Let the target speech and the interfering noise be represented by the expressions $s_m[n] \in \mathbb{R}$ and $b_m[n] \in \mathbb{R}$, respectively, for each microphone $m$, such that $x_m[n] = s_m[n] + b_m[n]$.
The STFT of the speech and noise interference corresponds to the expressions $S_m[l,k] \in \mathbb{C}$ and $B_m[l,k] \in \mathbb{C}$, respectively, and the Ideal Ratio Mask can be defined as:
\begin{equation}
    C[l,k] = \frac{\sum_{m=1}^{M}{|S_m[l,k]|^2}}{\sum_{m=1}^{M}{|S_m[l,k]|^2 + |B_m[l,k]|^2}},
\end{equation}
where $C[l,k] \in [0,1]$.

The objective is to estimate this IRM from the beamformed signal.
With two or more speakers active simultaneously, we need to address the permutation ambiguity to differentiate the target source from the interfering source(s).
We define the total power in the time-frequency domain as:
\begin{equation}
    |\hat{Y}[l,k]|^2 = \sum_{m=1}^{M}{|X_m[l,k]|^2}.
\end{equation}

As opposed to the power of the beamformed signal $|Y[l,k]|^2$ , the total power signal $|\hat{Y}[k,l]|^2$ ignores the phase information, and behaves as a reference signal without constructive or destructive interference.
The strategy here consists in using a deep neural network to estimate the IRM as follows:
\begin{equation}
    \hat{\mathbf{C}} = g_{DNN}\{\log(|\mathbf{Y}|^{\circ 2}), \log(|\hat{\mathbf{Y}}|^{\circ 2})\}
\end{equation}
where $\hat{\mathbf{C}} \in [0,1]^{L \times N/2+1}$ stands for the estimated IRM accross all $L$ frames and $N/2+1$ frequency bins.
The expressions $\log(|\mathbf{Y}|^{\circ 2})$ and $\log(\hat{|\mathbf{Y}}|^{\circ 2})$ consist of the spectrograms that contain the signals $\log|Y[l,k]|^2$ and $\log|\hat{Y}[l,k]|^2$ for each frame $l$ and frequency bin $k$.
A unidirectional two-layer Gated Recurrent Unit (GRU) neural network with $H=512$ hidden units architecture is proposed here.
The output of the GRU goes to a linear layer that maps to the number of frequency bins that match the frame size chosen for the STFT, and a sigmoid function as the IRM values lie in the interval $[0,1]$.
Note that the network needs to be causal to ensure minimum latency for this application, which justifies the unidirectional feature.
The network is trained by minimizing the mean-squared error (MSE) loss function weighted by the spectrograms:
\begin{equation}
loss = \left\lVert \mathbf{C} \odot \mathbf{|\mathbf{Y}|^{\circ 2}} - \hat{\mathbf{C}} \odot \mathbf{|\mathbf{Y}|^{\circ 2}} \right\lVert_2^2,
\end{equation}
where $\odot$ stands for the Hadamard product and $\lVert\dots\lVert_2^2$ corresponds to the $l^2$-norm.
This weighting gives more importance to time-frequency regions dominated by loud speech.

The estimated IRM can then be converted to a gain $G[l,k] \in [0,1]$ as follows:
\begin{equation}
    G[l,k] = \sqrt{\hat{C}[l,k]},
\end{equation}
where $\sqrt{\hat{C}[l,k]}$ stands for each time-frequency element in $\hat{\mathbf{C}}$.
The enhanced signal $Z[l,k]$ then corresponds to:
\begin{equation}
    Z[l,k] = G[l,k]Y[k,l].
\end{equation}

This can finally be converted back to the time-domain using an inverse STFT to generate $z[n] \in \mathbb{R}$.
This audio stream is then played back to the headphones as the enhanced signal.

\subsection{Hardware}

\begin{figure}[!ht]
    \centering
    \vspace{10pt}
    \includegraphics[width=\linewidth]{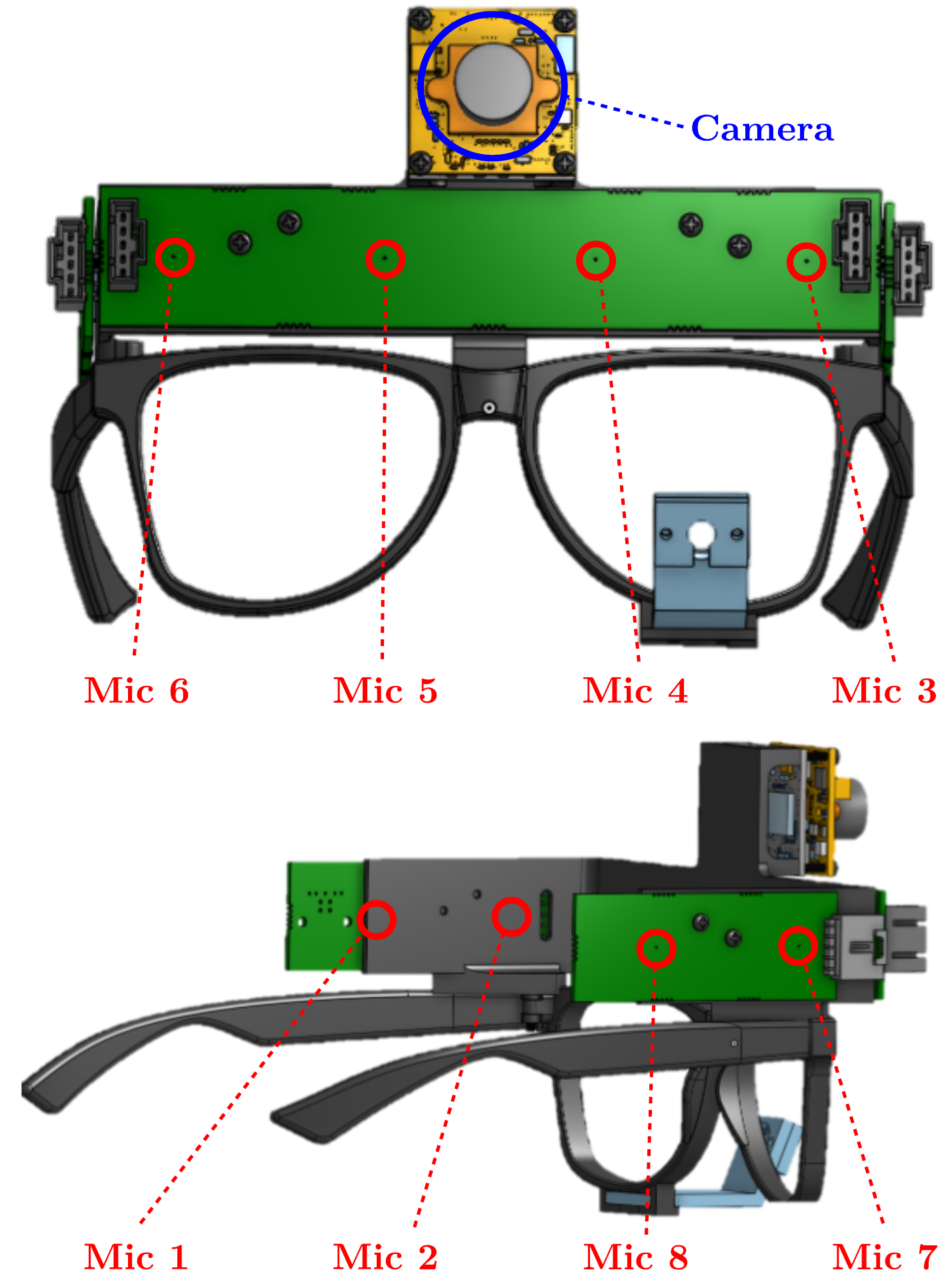}
    \caption{Computer-aided design of the 3-D printed eyeglasses with printed circuit boards and sensors}
    \label{fig:cad}
\end{figure}

A custom-made hardware is designed to provide audio feedback with minimum latency.
Audio acquisition is performed with eight MEMS microphones installed on printed circuit boards (PCBs) set on the front and sides of 3-D printed eye glasses, as shown in Fig. \ref{fig:cad}.

\begin{figure}[!ht]
    \centering
    \includegraphics[width=\linewidth]{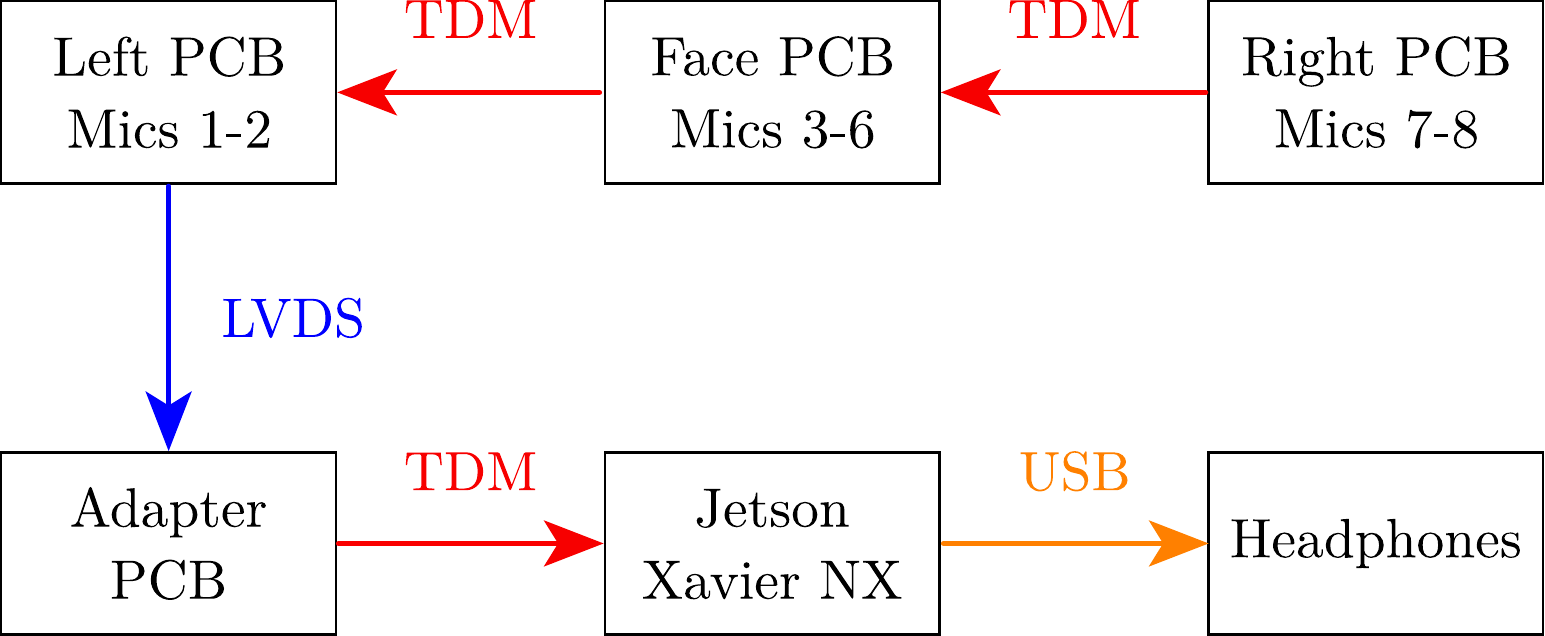}
    \caption{Connectivity between the different parts to route the audio signal from the microphones to the computer, and then the enhanced signal to the headphones}
    \label{fig:signal}
\end{figure}

Microphones are connected in daisy-chain using Time-Division Multiplexing (TDM) on the PCB.
Real-time processing is performed on a Jetson Xavier NX that can be installed in a backpack carried by the listener.
The high frequency signals therefore needs to be sent from the eyeglasses to the Jetson Xavier NX in the backpack, over a distance of approximately 1 m.
To make the signal robust to noise when sent through long wires, the PCB on the eyeglasses converts the digital audio signals to a Low Voltage Differential Signal (LVDS), which is received by another PCB mounted on the Jetson device.
The Jetson PCB converts the LVDS signals to TDM, and interfaces directly with the Jetson Xavier NX general input-output pins (I2S GPIOs, reconfigured to handle TDM).
As the audio GPIOs of Jetson are dedicated to the input with the TDM protocol, the processed output audio is forwarded via USB to a USB to AUX adapter, which is connected to the headphones.
Passive noise isolation (PNI) headphones are used, such that the user only hears the playback enhanced signal.
Figure \ref{fig:signal} summarizes the audio signal path from the microphones to the headphones.

The entire latency from the microphone to the playback signal (without processing) is measured to be 80 msec, while the algoritmic latency is 40 msec, for a total of 120 msec.
While this remains large for real-time enhancement (yet still under the 140 msec upper limit mentioned previously), the 80 msec latency could be reduced if the playback signal was generated with GPIOs instead of a USB output that converts the signal to an auxiliary signal sent to the headphones. 
This limitation currently comes from the fact that the Jetson Xavier NX is limited in terms of GPIOs for audio.

The camera can be interfaced via a standard USB port, connected directly to the Jetson Xavier NX.
The Jetson module also runs a server that provides a web page to which the user can connect using a smartphone via Wi-Fi.
This application provides the video stream captured by the camera mounted on the eyeglasses, and allows the user to select the face of the speaker to be tracked over time and whose speech needs to be enhanced.

\section{EXPERIMENTAL SETUP}
\label{sec:experimental_setup}

The proposed system is installed on a dummy head and sound sources are played individually to generate a training dataset in five different rooms.
The dummy head is positioned at three different positions in each room.
In four rooms, a loudspeaker plays speech (from the MS-SNSD dataset \cite{reddy2019scalable}) and is positioned at five random positions in the field of view of the camera, and a loudspeaker plays a noise source (also from the MS-SNSD dataset) and is positioned at thirteen different locations all around the dummy head.
In the fifth room, there are six random positions for the speech source and ten positions for the noise source, which are produced by participants instead of using a loudspeaker.
Each speech and noise source lasts ten seconds and is recorded individually, such that they can be mixed offline with different gains to create mixtures with SNRs in the interval $[0.5, 10]$ dB.
In total, speech and noise recordings are performed in $5$ rooms with different configurations (different microphone array, speech source and noise source positions) for a total of $560$ sec of recorded speech and $1920$ sec of recorded noise.
These recording are combined randomly to create scenarios with $1$ to $3$ interfering sources and random gains, for a total of $63389$ combinations of $10$ sec, which represents $176$ hours.

\begin{figure}[!ht]
    \centering
    \includegraphics[width=\linewidth]{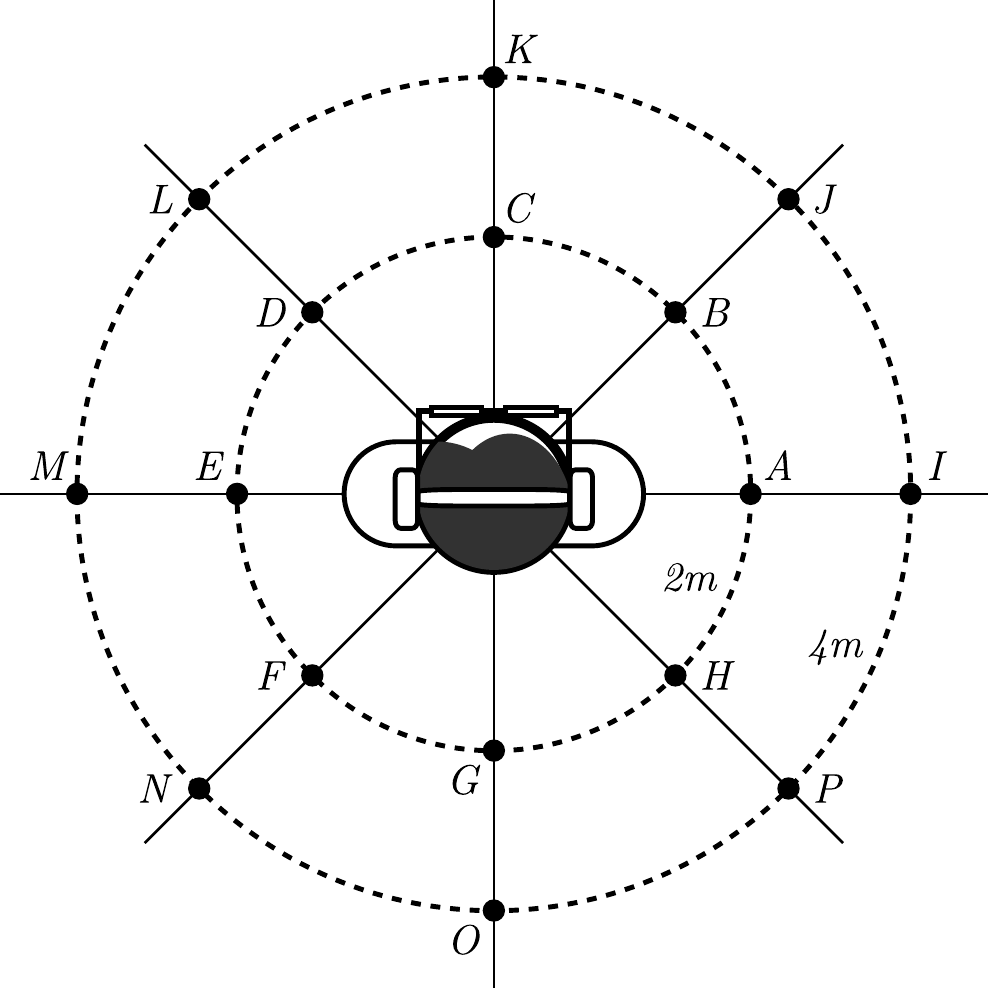}
    \caption{Experimental setup with sound source positions around the user}
    \label{fig:experimental_setup}
\end{figure}

For validation, sound sources are positioned at $16$ different positions around the dummy head, denoted by the letters $A$ to $P$ in figure \ref{fig:experimental_setup}.

A loudspeaker is positioned at a height of $165$ cm and plays five minutes of randomly chosen Librispeech \cite{panayotov2015librispeech} speech samples and MS-SNSD non-speech samples at each position ($A$ to $P$).
Different 10-second scenarios are then generated by mixing individual recordings, where the target sound source is located at a random position in the field of view of the system (positions $B$, $C$, $D$, $J$, $K$ or $L$), and the interfering sound source is located at a different position, this time in or out of the field of view.
When mixing sound sources, special care is taken to ensure the target and interfering source are at different position, and the sound segments correspond to different Librispeech samples.

\section{RESULTS AND DISCUSSION}
\label{sec:results}

\begin{figure*}
    \vspace{10pt}
    \centering
    \begin{subfigure}[b]{\linewidth}
    \centering
    \includegraphics[width=\textwidth]{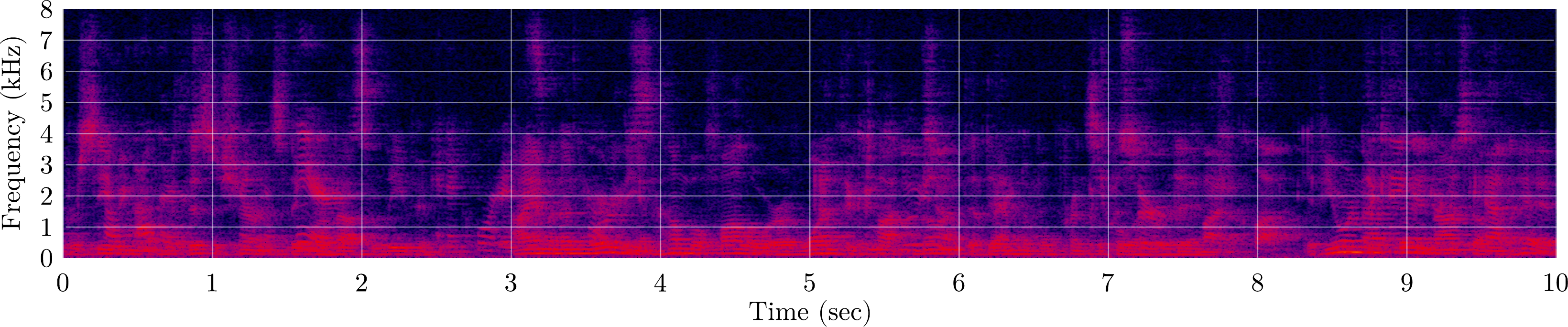}
    \caption{Input signal $\mathbf{X}_1$}
    \vspace{10pt}
    \end{subfigure}
    \begin{subfigure}[b]{\linewidth}
    \centering
    \includegraphics[width=\textwidth]{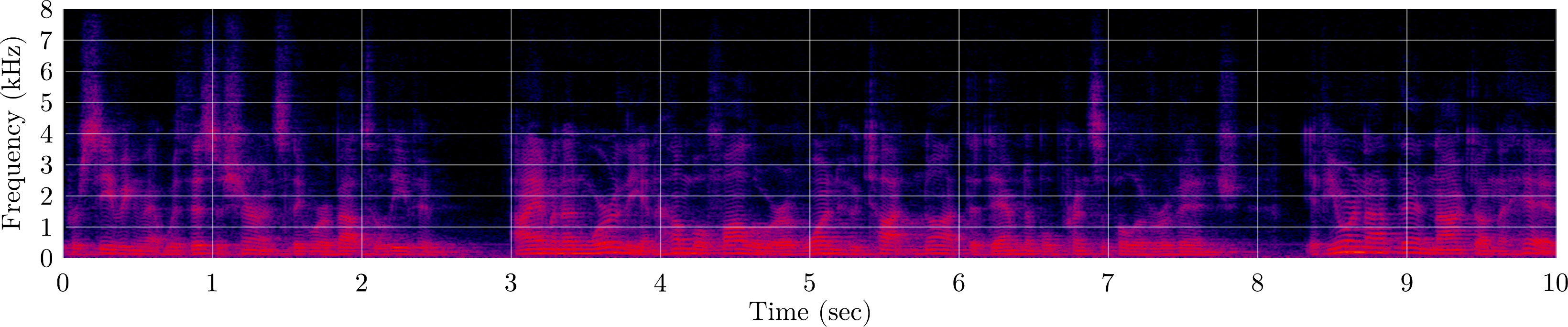}
    \caption{Target signal $\mathbf{S}_1$}
    \vspace{10pt}
    \end{subfigure}
    \begin{subfigure}[b]{\linewidth}
    \centering
    \includegraphics[width=\textwidth]{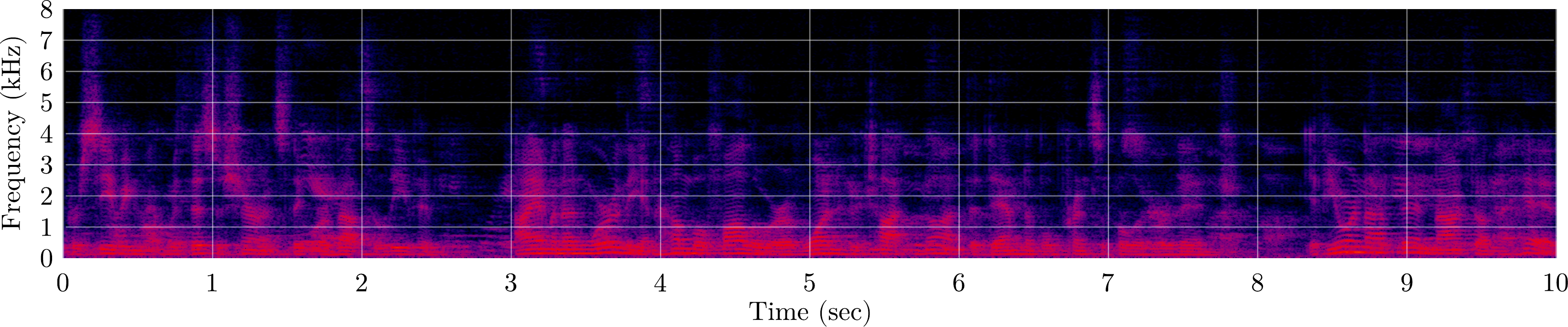}
    \caption{Postfiltered signal $\mathbf{Z}$}
    \vspace{10pt}
    \end{subfigure}
    \caption{Spectrograms of the input and processed signals. In this example, the SI-SDR of the input and postfiltered signals are 6.9 dB and 12.1 dB, respectively. The PESQ is 1.34 for the input signal, and 2.17 for the postfiltered signal. The STOI is 0.72 for the input and 0.88 for the postfiltered signal.}
    \label{fig:spex}
\end{figure*}

Figure \ref{fig:spex} shows the spectrograms of the input signal at microphone 1 ($\mathbf{X}_1$), the target signal only at microphone 1 ($\mathbf{S}_1$), and the postfiltered signal ($\mathbf{Z}$).
Some of the interfering phonemes are attenuated by beamforming and postfiltering.
There is still interference in the postfiltered spectrogram, suggesting that a beamformer with smaller sidelobes in the direction of the interfering sources and/or a more efficient postfiltering neural network could further improve the performances.

To evaluate the impact of the postfiltering step, we estimate Scale-Invariant Signal-to-Distortion Ratios (SI-SDRs), Perceptual Evaluation of Speech Quality (PESQ) and Short-Time Objective Intelligibility (STOI) values for the input and postfiltered signals in scenarios with speech interference (shown in Table \ref{tab:metrics_speech}) and speech and non-speech interference (shown in Table \ref{tab:metrics_nonspeech}).

\begin{table}[!ht]
    \centering
    \caption{Separation metrics for speech interference}
    \renewcommand{\arraystretch}{1.5}
    \begin{tabular}{c|cc}
    \hline
    \hline
        Metrics & Input & Postfiltered \\
    \hline
        SI-SDR (dB) & 0.70 & 2.39 \\
        PESQ & 1.28 & 1.43 \\
        STOI & 0.61 & 0.70 \\
    \hline
    \hline
    \end{tabular}
    \label{tab:metrics_speech}
\end{table}

\begin{table}[!ht]
    \centering
    \caption{Separation metrics for speech and non-speech interference}
    \renewcommand{\arraystretch}{1.5}
    \begin{tabular}{c|cc}
    \hline
    \hline
        Metrics & Input & Postfiltered \\
    \hline
        SI-SDR (dB) & 1.27 & 3.09 \\
        PESQ & 1.40 & 1.58 \\
        STOI & 0.59 & 0.69 \\
    \hline
    \hline
    \end{tabular}
    \label{tab:metrics_nonspeech}
\end{table}

Results demonstrate the system improves the speech quality, yet some further improvements could be made.
For instance, the GSS approach could be explored using the TDoA for each interfering speech sources, assuming each competing speaker is in the field of view of the camera.
The size of the dataset could also be increased to improve the postfiltering neural-network performances.
In fact, although there is a total of $176$ hours of mixture, they all come from the same 42 minutes of recorded signals.

\section{CONCLUSIONS}
\label{sec:conclusions}

In this paper, we demonstrate the feasibility of a complete real-time system for speech enhancement on embedded hardware.
The proposed approach uses face detection, tracking and verification to obtain the TDoA of the target source, which is then fed to a delay and sum beamformer.
The audio signal is then postfiltered with a gain predicted using a neural network.

The results suggest that, as future work, other types of beamformer should be explored (e.g. GSS) and more data could be collected to improve the performance of the neural network.
Different neural network architectures could also be explored for postfiltering (e.g. LSTM, ResNet, Transformers), as long as the latency is similar or smaller than the current network.
The current hardware and software could be used to validate these methods in real environments.
The hardware could also be revisited to use a direct output for the audio playback, to reduce the latency introduced by the USB link currently used.







\bibliographystyle{IEEEtran}
\bibliography{IEEEabrv, references}

\end{document}